\documentclass[%
 reprint,
 amsmath,amssymb,
 aps,
]{revtex4-2}
\global\long\def\dd{\mathrm{d}}%
\usepackage{graphicx}
\usepackage{xcolor}
\usepackage{dcolumn}
\usepackage{bm}

\begin{document}


\title{Black Hole shadows of $\alpha'$-corrected black holes}

\author{F. Agurto-Sepúlveda$^{1,2}$}
\author{J. Oliva$^1$}%
 \author{M. Oyarzo$^{1,3}$}%
 \author{D.R.G Schleicher$^2$}%
 
 \affiliation{%
 ${}^1$Departamento de Física, Universidad de Concepción, Casilla, 160-C, Concepción, Chile
 }%
 \affiliation{%
 ${}^2$Departamento de Astronomía, Universidad de Concepción, Casilla, 160-C, Concepción, Chile.
 }
 \affiliation{%
 ${}^3$Department of Applied Science and Technology, Politecnico di Torino,
C.so Duca degli Abruzzi, 24, I-10129 Torino, Italy
 }

\date{\today}

\begin{abstract}
In this paper we study the qualitative features induces by corrections to GR coming from String Theory, on the shadows of rotating black holes. We deal with the slowly rotating black hole solutions up to order $\mathcal{O}(a^3)$, to first order in $\alpha'$, including also the dilaton. We provide a detailed characterization of the geometry, as well as the innermost stable circular orbit (ISCO) and photon ring, and then we proceed to obtain the black hole images within the relativistic thin-disk model. We characterize the images by computing the diameter, displacement and asymmetry. A comparison with the Kerr case, indicates that all these quantities grow due to the $\alpha'$ correction, and that the departure from GR for different observable is enhanced depending on the angle of view, namely for the diameter the maximum departure is obtained when the system is face-on, while for the displacement and asymmetry the departure from GR is maximized for edge-on point of view. 
\end{abstract}

\maketitle


\section{Introduction}

Over the last decades, the detection of compact objects such as black holes has increased significantly, especially with new generations of instrumentation for the observation and detection of black holes. These go from interferometers that detect gravitational waves from black hole binaries such as the LIGO-VIRGO collaboration \cite{abbott2016observation}, to the direct observation of these objects with the Event Horizon Telescope (EHT). The latter has obtained the first images of supermassive black holes (SMBH) at the center of the M87 galaxy \cite{collaboration2019first}, as well as at the center of our own galaxy (Sgr A* SMBH)  \cite{akiyama2022first}. These results define historical and crucial achievements in testing the theory of General Relativity (GR) and serve as laboratories for testing it in the strong field regime. More recent results include the study of the relativistic jets of these systems, as well as the observation of polarization emission due to the magnetized environment \cite{event2021first,EHT:2023ujh,akiyama2024first}.

While GR has been able to describe the current observations to a high precision, it is well-known that there are both theoretical and experimental reasons that indicate the need of an extension of such framework, including the failure to predict the accelerating expansion of the Universe, to incorporate dark matter, as well as the incompatibility with quantum field theory. 

In this context, the inclusion of direct observation of SMBHs creates a great opportunity to test and constrain the effects of modified gravity theories in environments close to them, which will further improve with future facilities like the ngEHT (Next Generation Event Horizon Telescope) \footnote{The ngEHT web: https://www.ngeht.org/} project, that aims at increasing the baseline by incorporating satellites, thereby improving on the resolution within the images, allowing for instance for very precise measurements of the black hole spin \cite{ricarte2022ngeht}.

It is currently accepted that GR has to be interpreted as an effective field theory, emerging from a UV-complete framework, in the low energy limit. Within such setup, corrections to 
GR are expected, which come suppressed by a perturbative parameter. In the context of String Theory, such corrections come multiplied by $\alpha'$, namely by one over the string tension, and are in general hard to compute. Even more, since String Theory is consistently formulated in dimension ten, a compactification scheme is necessary to connect the theory with four-dimensional physics. The four-dimensional Einstein-Hilbert term, is supplemented by higher dimensional operators, which perturbativelly modify the dynamics and the structure of the solutions.

The aim of this work is to establish the qualitative features that such terms may imprint on the shadows of rotating black holes.

It is important to mention that the order of magnitude of $\alpha'$ is expected to be much lower than the values here considered, notwithstanding this paper is devoted to study in a qualitative manner the signals induced by the $\alpha'$-corrections on the black hole images. The paper is organized as follows: Section II introduces the theory on the string frame. It also contains the field equations that a perturbative solution may  fulfill, by expanding around a constant dilaton background and an Einstein geometry. Section III reviews the construction of the $\alpha'$-corrected, slowly rotating black hole up to $\mathcal{O}(a^3,a^2\alpha,\alpha^2)$, presenting the expressions for the temperature, entropy, angular velocity of the horizon and the first law of black hole thermodynamics. Section IV contains further properties of the spacetime that are relevant for the construction of the shadows. Section V, introduces the numerical model where the emission of the relativistic accretion disk is treated along the Novikov-Thorne model. The section also contains a review of the Ray-tracing code. Section VI presents the results of the simulations for the black hole shadows, for different values of $\alpha'$, focusing on the signals induced on the diameter, displacement and asymmetry of the images, induced by the stringy corrections. Section VII contains conclusion and further comments.

\section{The theory}

The theory we will consider emerges from the low-energy effective action of String
Theory including $\alpha^{\prime}$ corrections to the dilaton-graviton
sector. To order $\mathcal{O}(\alpha^{\prime2})$ terms, we are able
to perform a suitable field redefinition leading to field equations of second order for the metric and the dilaton \citep{Metsaev:1987zx}, \cite{Metsaev:1987bc} (see also 
\citep{Maeda:2011zn} and \citep{Agurto-Sepulveda:2022vvf})
\begin{align}
I[g_{\mu\nu},\phi] & =\int\dd^{4}x\sqrt{-g}e^{-2\phi}[R+4(\partial\phi)^{2} \nonumber\\
 & +\alpha(\mathcal{X}_{4}-16(\partial_{\mu}\phi\partial^{\mu}\phi)^{2}+\mathcal{O}(\alpha^{2})]\ ,\label{the action principle} 
\end{align}
where the Euler density is defined as
\begin{align}
\mathcal{X}_{4} & =R_{\mu\nu\rho\sigma}R^{\mu\nu\rho\sigma}-4R_{\mu\nu}R^{\mu\nu}+R^{2}\ .
\end{align}
Here $\alpha=\alpha'/8$. See \citep{Cano:2021rey} for the precise dimensional reduction and consistent truncation of String Theory leading to the action principle \eqref{the action principle} and \citep{Agurto-Sepulveda:2022vvf} for the field redefinition that allows connecting both scenarios. The field equations are
\begin{align}
R_{\mu\nu}+2\phi_{;\nu;\mu}+\alpha\mathcal{H}_{\mu\nu} & =0\,,\label{eom1}\\
R-4(\partial\phi)^{2}+\alpha\mathcal{P} & =0\,,\label{eom2}
\end{align}
where
\begin{align}
\mathcal{H}_{\mu\nu} & =-4\phi_{;\nu;\mu}(\partial\phi)^{2}+8\phi^{;\sigma}{}_{;\mu}\phi_{;\nu;\sigma}\\
 & -16\phi_{(;\mu}\phi_{;\nu);\sigma}\phi^{;\sigma}+2(\phi^{;\sigma;\lambda}-2\phi^{;\sigma}\phi^{;\lambda})R_{\mu\sigma\lambda\nu}\nonumber \\
 & +g_{\mu\nu}[-56(\partial\phi)^{4}-12\phi_{;\lambda;\sigma}\phi^{;\lambda;\sigma}\nonumber \\
 & -24\phi_{;\rho;\sigma}\phi^{;\rho}\phi^{;\sigma}+\frac{1}{2}R_{\lambda\delta\rho\sigma}R^{\lambda\delta\rho\sigma}]\,,\nonumber \\
\mathcal{P} & =R_{\lambda\delta\rho\sigma}R^{\lambda\delta\rho\sigma}-16\phi_{;\mu;\nu}\phi^{;\mu;\nu}+\\
 & -64\phi^{;\rho;\sigma}\phi_{;\rho}\phi_{;\sigma}-96(\partial\phi)^{4}\,.\nonumber 
\end{align}

Since the action (\ref{the action principle}) is
consistent neglecting $\mathcal{O}(\alpha^{2})$, we are interested
in studying $\alpha^{\prime}$--corrected configurations
$\overline{g}_{\mu\nu}$ and $\overline{\phi}$ satisfying the equations
(\ref{eom1}) and (\ref{eom2}) with $\alpha=0$. Therefore, the general
correction should be of the form
\begin{align}
g_{\mu\nu} & =\overline{g}_{\mu\nu}+\alpha\gamma_{\mu\nu}+\mathcal{O}(\alpha^{2})\,,\\
\phi & =\overline{\phi}+\alpha\tau+\mathcal{O}(\alpha^{2})\,.
\end{align}
where $\gamma_{\mu\nu}$ is a symmetric perturbative tensor and $\tau$
is a scalar function of the spacetime coordinates. The situation in
which $\overline{\phi}=\phi_{0}$ is a constant is particularly interesting
since the equations for $\overline{g}_{\mu\nu}$ consistently reduce to the vacuum
Einstein equations, and we will focus in such case through out
this work. It is interesting to notice that due to the fact that
the dilaton appears linearly in the equation (\ref{eom1}), then a
consistent correction at order $\alpha$ to the metric is allowed.
Consider the leading term of the dilaton being a constant and the
leading term of the metric a solution of the vacuum Einstein equations,
then a huge simplification of the equations (\ref{eom1}) and (\ref{eom2})
takes place leading to
\begin{align}
\alpha\Pi_{\mu\nu}+2\alpha\overline{\nabla}_{\mu}\partial_{\nu}\tau+\frac{1}{2}\alpha\overline{g}_{\mu\nu}\overline{\mathcal{K}} & =0\,,\label{phi const  eom1}\\
\alpha\overline{\Pi}_{\mu}{}^{\mu}+4\alpha\overline{\nabla}^{\mu}\partial_{\mu}\tau+\alpha\overline{\mathcal{K}} & =0\,.\label{phi const eom2}
\end{align}
where
\begin{align*}
\Pi_{\mu\nu} & =\overline{\nabla}^{\lambda}\overline{\nabla}_{(\mu}\gamma_{\nu)\lambda}-\frac{1}{2}\overline{\nabla}^{\lambda}\overline{\nabla}_{\lambda}\gamma_{\mu\nu}-\frac{1}{2}\overline{\nabla}_{\mu}\partial_{\nu}\gamma_{\lambda}{}^{\lambda}\,,\\
\overline{\mathcal{K}} & =R_{\lambda\delta\rho\sigma}R^{\lambda\delta\rho\sigma}|_{\overline{g}_{\mu\nu}}\,.
\end{align*}
The barred covariant derivatives are computed with the Levi-Civita
connection associated to the metric $\overline{g}_{\mu\nu}$. This
equation can be integrated in a variety of situations as was shown
in \citep{Agurto-Sepulveda:2022vvf} for accelerating black holes
and slowly rotating black holes. We will analyze in detail the later
situation in which the seed metric is the Kerr spacetime expanded up to second order on the rotation parameter $a=GJ/M$.
\section{Slowly rotating $\alpha' -$ corrected BH}

The slowly rotating black hole solution found in \citep{Agurto-Sepulveda:2022vvf}
corresponds to a consistent solution of the equations (\ref{phi const  eom1})
and (\ref{phi const eom2}) for $\overline{g}_{\mu\nu}$ being the
slowly rotating Kerr metric with rotation parameter $a$, expanded up to $a^{2}$ neglecting terms $\mathcal{O}(a^{3})$, and extending to the stationary case the early found static solutions in $\alpha'$-corrected GR in \cite{Callan:1988hs}. We find that
it is possible to integrate the equation even considering the logarithmic
branch, controlled by an integration constant $b$, in the $\alpha$-corrected
static configuration. An analytic solution of (\ref{phi const  eom1})-(\ref{phi const eom2})
at non-linear level in the rotation parameter is still not known in
the literature, nevertheless higher order correction in $a$ could be obtained along the lines of \citep{Cano:2021rey}, also in Einstein theory in the presence of a Gauss-Bonnet term in arbitrary dimension \cite{Konoplya:2020fbx}. The slowly rotating spacetime, neglecting orders $\mathcal{O}(a^{2}\alpha,a^{3},\alpha^2)$, 
including the logarithmic branch is given by
\begin{align}
g_{tt} & =-1+\frac{2M}{r}+\frac{2Ma^{2}\cos^{2}\theta}{r^{3}}-\alpha f_{1}(r)\,,\label{alpha-metric}\\
g_{t\phi} & =-\frac{2Ma\sin^{2}\theta}{r}+a\alpha h_{1}(r,\theta)\,,\\
g_{rr} & =\frac{1}{1-\frac{2M}{r}+\alpha g_{1}(r)}-a^{2}\frac{(2M-r)\cos^{2}\theta+r}{(2M-r)^{2}r}\,,\\
g_{\theta\theta} & =r^{2}+a^{2}\cos\theta\,,\\
g_{\phi\phi} & =(r^{2}+a^{2}+\frac{2Ma^{2}\sin^{2}\theta}{r})\sin^{2}\theta\,,
\end{align}
where
\begin{align}
f_{1}(r) & =\frac{8M^{2}}{r^{4}}+\frac{10M}{3r^{3}}+\frac{4}{r^{2}}-\frac{4}{Mr}-\frac{4bM}{r}+\\
 & +\frac{b(r-3M)}{r}\log\left[\left(1-\frac{2M}{r}\right)^{2}\right]\nonumber \\
g_{1}(r) & =-\frac{40M^{2}}{3r^{4}}+\frac{2M}{r^{3}}+\frac{2}{r^{2}}+\\
 & -\frac{bM}{r}\log\left[\left(1-\frac{2M}{r}\right)^{2}\right]\,,\nonumber \\
h_{1}(r,\theta) & =\left(\frac{8M^{2}}{r^{4}}+\frac{6M}{r^{3}}+\frac{6}{r^{2}}+\right.\\
 & \left.-\frac{3bM}{r}\log\left[\left(1-\frac{2M}{r}\right)^{2}\right]\right)\sin^{2}\theta\,.
\end{align}
While the dilaton field is given by
\begin{align*}
\phi & =\phi_{0}+\alpha\left(-\frac{4M}{3r^{3}}-\frac{1}{r^{2}}-\frac{1}{Mr}+\frac{b}{2}\log\left[\left(1-\frac{2M}{r}\right)^{2}\right]\right)
\end{align*}
In spite of the generalization including the logarithmic branch to
the slowly rotating solution, in this work we will concentrate in the $b=0$ case, since only such case has a well-defined static black hole limit. When $b\neq 0$, the static configuration has metric functions $g_{tt}$ and $g^{rr}$ that vanish at different points, and therefore the spacetime lacks of an event horizon in such case. Requiring $g_{tt}$ and $g^{rr}$ to vanish at the same point, up to $\mathcal{O}\left(\alpha^2\right)$ requires setting $b=0$ (see \citep{Agurto-Sepulveda:2022vvf}).  Using the Iyer-Wald method to compute
conserved charges for the theory (\ref{the action principle}) as
was explained in \citep{Agurto-Sepulveda:2022vvf} the mass and the
angular momentum neglecting $\mathcal{O}(a^{3},a^{2}\alpha)$ are
respectively
\begin{align}
\mathcal{M} & =\frac{M}{G}+\frac{\alpha}{GM}\,,\\
\mathcal{J} & =\frac{aM}{G}\,.
\end{align}

The deviation tensor $\gamma_{\mu\nu}$ is present in the components
$g_{tt},g_{t\phi}$ and $g_{rr}$. Note also that the $\gamma_{\mu\nu}$
goes to zero fast enough as $r\to\infty$ thus the spacetime is asymptotically
flat, and the Iyer-Wald charges converge \footnote{See also \cite{Caceres:2023gfa} for a finite action principle in the context of Horndeski theories of the form \eqref{the action principle} that leads to finite charges via holographic renormalization in AdS.}. The Hawking temperature and entropy, computed from the
Wald formula \citep{Wald:1993nt}, of the slowly rotating black hole (\ref{alpha-metric})
with $b=0$ do not receive an $a\alpha$ correction and are given
by
\begin{align}
T & =\frac{1}{8\pi M}-\frac{\alpha}{8\pi M^{3}}-\frac{a^{2}}{32\pi M^{3}}\,,\\
\mathcal{S} & =\frac{4M^{2}\pi}{G}+\frac{12\pi\alpha}{G}-\frac{\pi a^{2}}{G}\,,
\end{align}
where consistently with the scenario we are dealing with, we have neglected terms $\mathcal{O}(a^{3},a^{2}\alpha)$. The angular
velocity of the horizon of the slowly rotating black hole is
\begin{align}
\Omega_{\mathrm{bh}} & =\frac{a}{4M^{2}}-\frac{3a\alpha}{4M^{4}}+\mathcal{O}(a^{3},a\alpha^2)\,.
\end{align}
The first law of thermodynamics holds by considering the variation
with respect to the integration constants $M$ and $a$ and considering
the variation of the perturbative parameter $\delta a$ being of order
$a$ 
\begin{align}
\delta\mathcal{M} & =T\delta\mathcal{S}+\Omega_{\mathrm{bh}}\delta\mathcal{J}\,,
\end{align}
again neglecting $\mathcal{O}(a^{3},a^{2}\alpha,\alpha^{2})$.

\section{Spacetime properties}
In this section we will describe further properties of the slowly rotating $\alpha'$-corrected geometry, comparing them with the Kerr slowly rotating case. These properties will be important for the simulations of the black hole shadows.

\label{sec:headings}

\subsection{Properties of the black hole geometry}

To determine the Petrov type of the slowly rotating spacetime we choose the following orthonormal tetrad
\begin{align}
e^{0} & =\sqrt{-g_{tt}}\dd t-\frac{g_{t\phi}}{\sqrt{-g_{tt}}}\dd\phi\,,\\
e^{1} & =\sqrt{g_{rr}}\dd r\,,\quad e^{2}=\sqrt{g_{\theta\theta}}\dd\theta\,,\nonumber \\
e^{3} & =\sqrt{g_{\phi\phi}-\frac{g_{t\phi}^{2}}{g_{tt}}}\dd\phi\,,\nonumber 
\end{align}
which must be expanded in $a$ and $\alpha$ neglecting $\mathcal{O}(a^{3},\alpha^{2},a^{2}\alpha)$.
The null tetrad reads
\begin{align}
k_{\mu} & =\frac{1}{\sqrt{2}}(e^{0}{}_{\mu}+e^{3}{}_{\mu})\,,\,l_{\mu}=\frac{1}{\sqrt{2}}(e^{0}{}_{\mu}-e^{3}{}_{\mu})\,,\\
m_{\mu} & =\frac{1}{\sqrt{2}}(e^{1}{}_{\mu}-ie^{2}{}_{\mu})\,,\,\overline{m}_{\mu}=\frac{1}{\sqrt{2}}(e^{1}{}_{\mu}+ie^{2}{}_{\mu})\,.\nonumber 
\end{align}
The components of the Weyl tensor in the null basis are 
\begin{align}
 & \Psi_{0}=C_{\mu\nu\rho\sigma}k^{\mu}m^{\nu}k^{\rho}m^{\sigma}\,,\,\Psi_{1}=C_{\mu\nu\rho\sigma}k^{\mu}l^{\nu}k^{\rho}m^{\sigma}\,,\nonumber \\
 & \Psi_{2}=C_{\mu\nu\rho\sigma}k^{\mu}m^{\nu}\overline{m}^{\rho}l^{\sigma}\,,\,\Psi_{3}=C_{\mu\nu\rho\sigma}l^{\mu}k^{\nu}l^{\rho}\overline{m}^{\sigma}\,,\nonumber \\
 & \Psi_{4}=C_{\mu\nu\rho\sigma}l^{\mu}\overline{m}^{\nu}l^{\rho}\overline{m}^{\sigma}\,.
\end{align}
For the spacetime (\ref{alpha-metric}) the following expressions hold
\begin{align*}
\Psi_{0} & =\mathcal{O}(a^{3},\alpha^{2},a^{2}\alpha)\,,\\
\Psi_{3} & =\mathcal{O}(a^{3},\alpha^{2},a^{2}\alpha)\,,\\
\Psi_{0}\Psi_{4}-9(\Psi_{2})^{2} & =\mathcal{O}(a^{3},\alpha^{2},a^{2}\alpha)\,.
\end{align*}
Indicating that the corrected spacetime is Petrov type D, as the seed
spacetime.

The event horizon of the static black hole is located at
\begin{equation}
r_{+} =  2M + \dfrac{\alpha}{6M}\ ,
\end{equation}
which covers a curvature singularity located at $r=0$
\begin{eqnarray} \label{Kretsch-alpha}
    R_{ \mu \nu \rho \sigma} R^{ \mu \nu \rho \sigma} &=& \frac{48 M^{2}}{r^{6}} -\frac{1008 \cos^{2} \theta M^{2} a^{2}}{r^{8}} \\
    &+&\frac{\left(-128 M^{3}-96 M^{2} r -96 M \,r^{2}+96 r^{3}\right) \alpha}{r^{9}}.\nonumber
\end{eqnarray}
The rotation induces a modification of the location of the event horizon, leading to
\begin{equation}
    r_{+} = 2M - \frac{a^2}{2M}  + \frac{\alpha}{6M} + \mathcal{O}(a^2 \alpha, \alpha^{2})\ ,
\end{equation}
while the ergosphere is located at
\begin{equation}
    r_{erg}^{\alpha'} = 2M - \frac{a^2}{2M} \cos^{2} \theta - \dfrac{\alpha}{12M^{2}} + \mathcal{O}(a^2 \alpha, \alpha^2).
\end{equation}
 Note that the $\alpha^\prime$-corrected ergosphere is smaller than the ergosphere of Kerr, namely  $r^{\alpha^\prime}_{\mathrm{erg}}<r_{\mathrm{erg}}^{SK}$, while the location of the event horizon of the $\alpha^\prime$-corrected geometry is larger than the location of the Kerr event horizon.

We will be exploring the geodesic orbits on this geometry, therefore, we impose the preservation of the Lorentzian signature as a consistency constraint on the perturbative construction. Since, the determinant of the Slowly rotating Kerr metric is
\begin{equation}
    g^{SK} = -r^{2} \sin^{2} \theta   \left(2 a^{2} \cos^{2 }\theta +r^{2}\right)\ ,
\end{equation}
the determinant of the spacetime defined by the metric functions (\ref{alpha-metric}) is
\begin{eqnarray}
    g^{\alpha'} =  g^{SK} + \frac{\alpha  \left(32 M^{2}r+18 M r^2 +12 r^{3}\right)  \sin^{2} \theta}{3 M}.
\end{eqnarray}
 As we remark, the $\alpha^{\prime}$-correction in the metric tensor goes to zero at spatial infinity, thus the signature of the metric can change only in the interior of the spacetime. In particular we are interested in the signature of the metric in the near horizon region. To first order in $\alpha$, the quotient of the respective determinants reads
\begin{eqnarray}
    \frac{g^{\alpha'}}{g^{SK}} = 1 - \alpha \left(\frac{32M}{3r^{3}} + \frac{6}{r^2} + \frac{4}{Mr}\right) \, . \label{ration galpha gSK}
\end{eqnarray}
To preserve the signature of the spacetime outside the horizon, namely $r\geq r_+$, it is a necessary condition that the second term of \eqref{ration galpha gSK} to be smaller than 1. Close to the horizon $r=r_{+}^{\alpha'}$, the signature should be preserved as well
\begin{equation}
    \frac{g^{\alpha'}}{g^{SK}} (r_{+}) = 1 - \frac{29 \alpha}{6M^2} + \mathcal{O}(a^2 \alpha,\alpha^2),
\end{equation}
implying
\begin{equation} \label{upper-limit}
    \alpha < \frac{6M^2}{29} + \mathcal{O}(a^{2}\alpha, \alpha^{2}),
\end{equation}
which is a consistent upper limit for the $\alpha$ coupling.

The circular geodesic motion around compact objects for a stationary axisymmetric space-time can be described with the equatorial approximation, $|\theta -\frac{\pi}{2}| \ll 1$. The geodesic equation
\begin{equation}\label{geq}
    \frac{d^{2}x^{\alpha}}{d\tau^{2}} = - \Gamma^{\alpha}_{\beta} {}_{\gamma} \frac{dx^{\beta}}{d\tau}\frac{dx^{\gamma}}{d\tau} ,
\end{equation}
along the $r$ component reduces to 
\begin{equation}
    \frac{1}{2}\partial_{r} g_{tt} \left(\frac{dt}{d \tau}\right)^{2} + \partial_{r} g_{t\phi} \frac{d t}{d\tau} \frac{d\phi}{d\tau}+  \frac{1}{2} \partial_{r} g_{\phi \phi} \left(\frac{d\phi}{d\tau}\right)^{2}= 0.
\end{equation}
From where one obtains the Keplerian frequency in the equatorial plane

\begin{eqnarray}
    \label{Keplerian-frequency}
    \Omega_{\phi} &=& \frac{p^{\phi}}{p^{t}} =\frac{-\partial_{r} g_{t\phi} \pm \sqrt{(\partial_{r}g_{t\phi})^{2} - \partial_{r}g_{tt} \partial_{r} g_{\phi \phi}}}{\partial_{r} g_{\phi \phi}},\\
    p^{\phi}&\equiv& \dot{\phi}\, ,\quad p^{t}\equiv \dot{t}\, ,
\end{eqnarray}
where the dots imply derivative w.r.t. the affine parameter, and the upper sign refers to prograde orbits and the lower sign to retrograde orbits, depending on the value of the spin parameter $a$.

Since the metric $g_{\mu \nu}$ is stationary and axisymmetric, it has a timelike and an azimuthal Killing vector, leading to the existence of two conserved quantities, the specific energy $\tilde{E}$ and the specific angular momentum $\tilde{L}$. Those correspond to the $p_t = -\tilde{E}$ and $p_{\phi} = \tilde{L}$ components of the four-momentum, from which the following geodesic equations can be derived, 

\begin{equation}
    \dot{t} = \frac{\tilde{E} g_{\phi \phi} + \tilde{L}g_{t \phi}}{g_{t \phi}^{2} - g_{t t} g_{\phi \phi}}, \\ 
\end{equation}
\begin{equation}
    \dot{\phi} = - \frac{\tilde{E} g_{t \phi} + \tilde{L}g_{t t}}{g_{t \phi}^{2} - g_{t t} g_{\phi \phi}}.
\end{equation}

The four momentum is normalized as $p^{\mu} p_{\mu}= -\epsilon^{2}$, 
where $\epsilon^{2} =+1,-1,$ for time-like and space-like geodesics, respectively. As usual, it is possible to obtain the expression for the effective potential in the equatorial plane as follows
\begin{equation}
    g_{rr} \dot{r}^{2} = V_{eff}(r)\ ,
\end{equation}
where the effective potential is given by
\begin{equation}
    V_{eff}(r) = - \epsilon^{2} + \frac{\tilde{E}^{2} g_{\phi \phi} + 2\tilde{E}\tilde{L} g_{t \phi} + \tilde{L}^{2}g_{t t}}{g_{t \phi}^{2} - g_{t t} g_{\phi \phi}}\ .\label{generic-potential}
\end{equation}
Our primary focus is on circular orbits confined to the equatorial plane for massive, timelike particles ($\epsilon^{2} = 1$), for which the orbits are governed by $V_{eff}(r) = 0=\frac{\partial}{\partial r} V_{eff}(r)$.

These conditions lead to
\begin{equation}\label{Energy}
    \tilde{E} = \frac{- g_{tt} - g_{t\phi}\Omega_{\phi}}{\sqrt{-g_{tt} - 2 g_{t\phi}\Omega - g_{\phi \phi}\Omega^{2}}},
\end{equation}
\begin{equation}\label{Angular-Momentum}
    \tilde{L} =  \frac{ g_{t\phi} + g_{\phi \phi}\Omega_{\phi}}{\sqrt{-g_{tt} -2g_{t\phi}\Omega - g_{\phi \phi}\Omega^{2}}}.
\end{equation}
On the other hand, since one of our goals is to obtain simulations of a thin accretion disk around a black hole described by this solution, we must determine the location of ISCO. It is possible to find it from the condition $\partial^{2} V_{eff}(r)/\partial r^{2}=0$, which leaves us with the following expression
\begin{equation}\label{ISCO-eq1}
    \tilde{E}^{2} \partial_{r}^{2} g_{\phi \phi} + 2\tilde{E}\tilde{L} \partial_{r}^{2} g_{t \phi} + \tilde{L}^{2} \partial_{r}^{2} g_{t t} - \partial_{r}^{2} (g_{t \phi}^{2} - g_{t t} g_{\phi \phi}) =0.
\end{equation}

Since the metric given by Eq. (\ref{alpha-metric}) satisfies all the characteristics previously mentioned, it is possible to obtain the Keplerian frequency from Eq. (\ref{Keplerian-frequency}) in $\mathcal{O}(\alpha)$, leading to
\begin{eqnarray}
    \Omega_{\phi} &=&  -\frac{Ma}{r^{3}} \pm \sqrt{\frac{M}{r}}\frac{ \left(M^{3} a^{2}+M^{2} r^{3}\right)}{r^{4} M^{2}}\notag \\ 
    &\pm&\notag  \frac{\alpha \left(-8 M^{3}-\frac{5}{2} r \,M^{2}-2 M \,r^{2}+r^{3}\right)}{r^{\frac{9}{2}} M^{\frac{3}{2}}}\\ &+&  \frac{a \alpha\left(32 M^{2}+18 M r +12 r^{2}\right) }{2 r^{6}}\ .\label{Keplerian-frequencyfinal}
\end{eqnarray}
The expressions for the specific energy and angular momentum, from Eqs. (\ref{Energy}-\ref{Angular-Momentum}) at $\mathcal{O}(\alpha)$ are
\begin{align}\label{Specific-Energy}
\tilde{E}= & \frac{r-2M}{r^{1/2}(r-3M)^{1/2}}+\frac{a^{2}M^{2}}{2\sqrt{r}(r-3M)^{5/2}}\\ \notag
 & \mp \frac{aM^{3/2}}{r(r-3M)^{3/2}}\\ \notag
 & -\alpha\frac{(34M^{3}+29M^{2}r-36Mr^{2}+6r^{3})}{6Mr^{5/2}(r-3M)^{3/2}}\\ \notag
 & \mp \frac{\alpha a}{6\sqrt{M}r^{4}(r-3M)^{5/2}}(120M^{4}-16M^{3}r+\\
 & +15M^{2}r^{2}-12Mr^{3}+6r^{4}) \notag
\end{align}
and 
\begin{align}
\tilde{L}= \label{Specific-Angular}& \pm \frac{rM^{1/2}}{(r-3M)^{1/2}}-a\frac{3M(r-2M)}{\sqrt{r}(r-3M)^{3/2}}+\\ \notag
 & \pm \frac{a^{2}M^{1/2}}{2r(r-3M)^{5/2}}(6M^{2}-5Mr+2r^{2})+\\ \notag
 & +\frac{\alpha a}{2r^{5/2}(r-3M)^{5/2}}(18M^{3}+25M^{2}r\\ \notag
 & -6Mr^{2}-6r^{3})+\\ \notag
 & \pm \frac{\alpha}{6r^{2}M^{3/2}(r-3M)^{3/2}}(72M^{4}-28M^{3}r\\
 & -3M^{2}r^{2}-12Mr^{3}+6r^{4})\notag
\end{align}
where the upper and lower sign correspond to prograde and retrograde orbits, respectively, while the effective potential takes the form 
\begin{align} \label{effective-potential}
     V_{eff} (r) = & \left(r^{3} + \left(2 M +r \right) a^{2}\right) E^2\\ \notag
    +& \left(\frac{\left(32 M^{2} r^{3}+18 r^{4} M +12 r^{5}\right) \alpha}{3 M \,r^{3}}\right) E^{2} \\ \notag
    +&\left(-\frac{4 \left(20 M^{2}+9 M r +3 r^{2}\right) a \alpha}{3 r^{3}}-4 M a \right) L E  \\ \notag
    +&\left(-r +2 M +\frac{\left(40 M^{3}-6 r \,M^{2}-6 r^{2} M \right) \alpha}{3 M \,r^{3}}\right) L^{2}\\ 
    -&r^{3}+2 r^{2} M -a^{2} r +\frac{\left(40 M^{3} r^{2}-6 M^{2} r^{3}-6 r^{4} M \right) \alpha}{3 M \,r^{3}}.\notag
\end{align}

Consequently, from Eq. (\ref{ISCO-eq1}) the expression for the ISCO is

\begin{eqnarray}\label{ISCO-eq2}
    &&r_{isco}^{\alpha'} = r_{isco}^{SK} +   \alpha \, \frac{163}{54 M} 
    - a \alpha \frac{362\, \sqrt{6} }{243 M^{2}} + \mathcal{O}(a^{2}\alpha,\alpha^{2})\, , 
\end{eqnarray}
where 
\begin{equation}
    r_{isco}^{SK} = M\left(6 -\frac{4 \sqrt{6} a}{3M} -\frac{7 a^{2}}{18 M^{2}}\right) + \mathcal{O}(a^3).
\end{equation}

\begin{figure}
	\centering
	\includegraphics[scale=0.5]{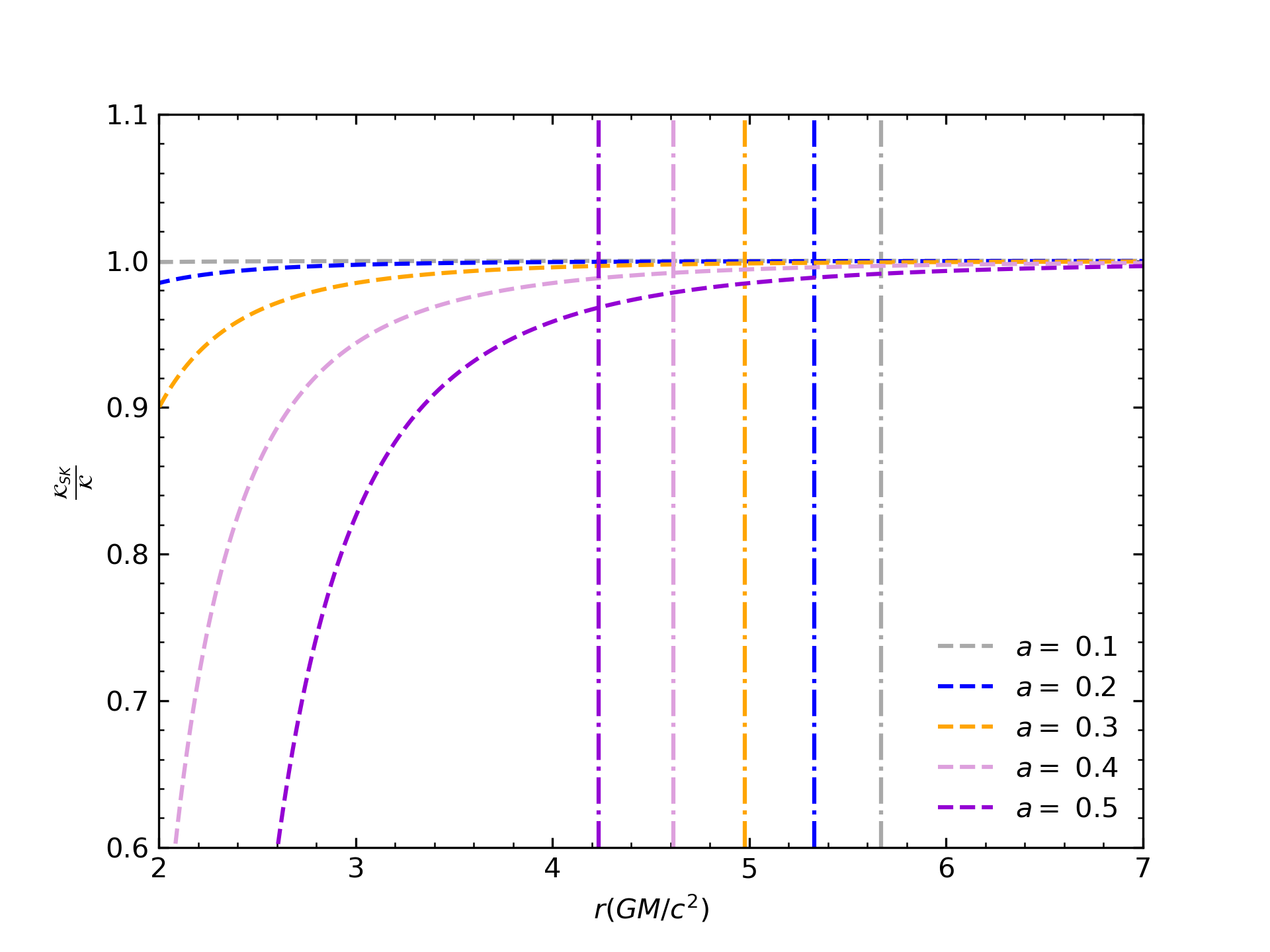}
        \caption{Ratio between the Kretschmann scalar in the slowly rotating Kerr metric and the standard Kerr metric, for different values of the spin parameter $a$, where vertical lines are the location of the ISCO for each $a$.}
	\label{Kretsch-limit}
\end{figure}

We use the Kretschmann scalar to define an upper limit for the rotation parameter $a$ to study the motion of test particles and the accretion disks around slowly rotating space-times. Figure \ref{Kretsch-limit} shows the ratio between the Kretschmann scalar for the slowly rotating approximation,
\begin{equation}
    \mathcal{K}_{SK} = \frac{48 M^{2}}{r^{6}} -\frac{1008 \cos^{2} \theta M^{2} a^{2}}{r^{8}} +\mathcal{O}(a^{3}),
\end{equation}
and the Kretschmann scalar for the regular Kerr solution,
\begin{eqnarray}
    \mathcal{K}&=& \frac{-48 M^{2}}{\left(r^{2}+a^{2} \cos^{2} \theta\right)^{6}} \Big(a^{6} \cos^{6} \theta -15 \cos^{4}\theta a^{4} r^{2} \nonumber \\ 
    &+& 15 \cos^{2}\theta  a^{2} r^{4}-r^{6}\Big),
\end{eqnarray}
which near the ISCO, starts departing from one for values around $a \approx 0.3M$, in consequence we restrict our attention to values of the spin parameter $a \lesssim 0.3M$.

\begin{figure*}
	\centering
	\includegraphics[scale=0.5]{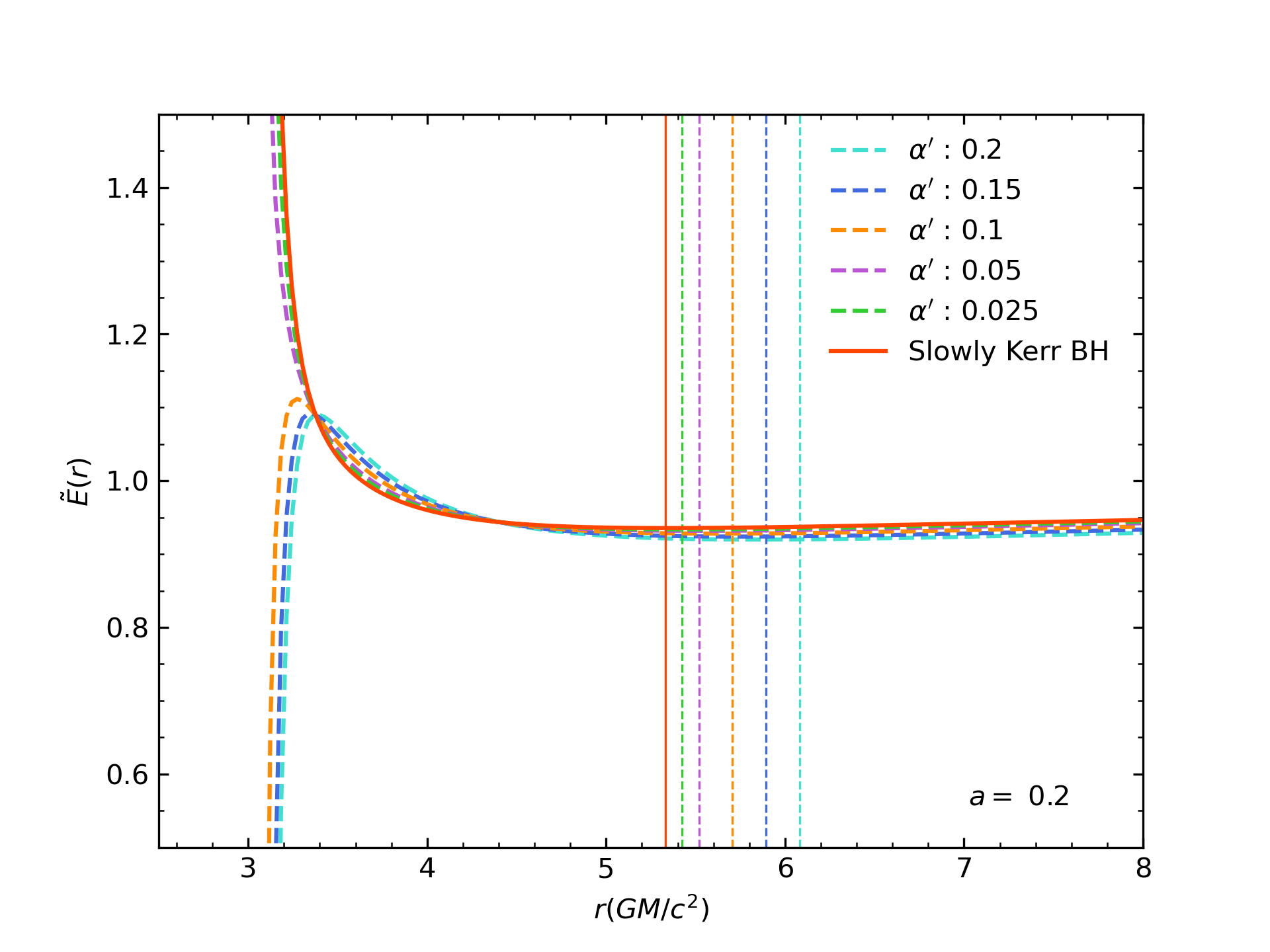}
        \includegraphics[scale=0.5]{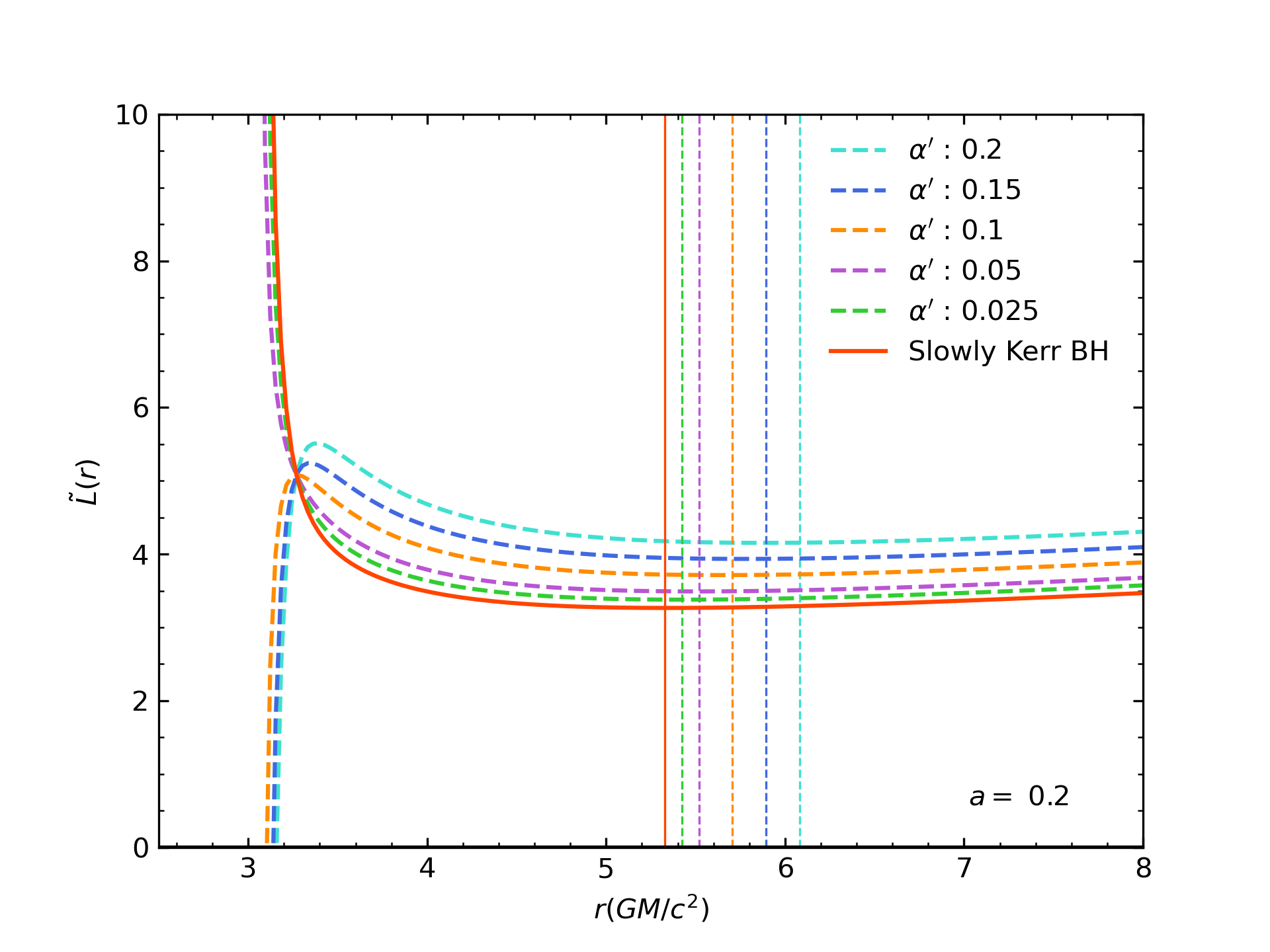}
        \includegraphics[scale=0.5]{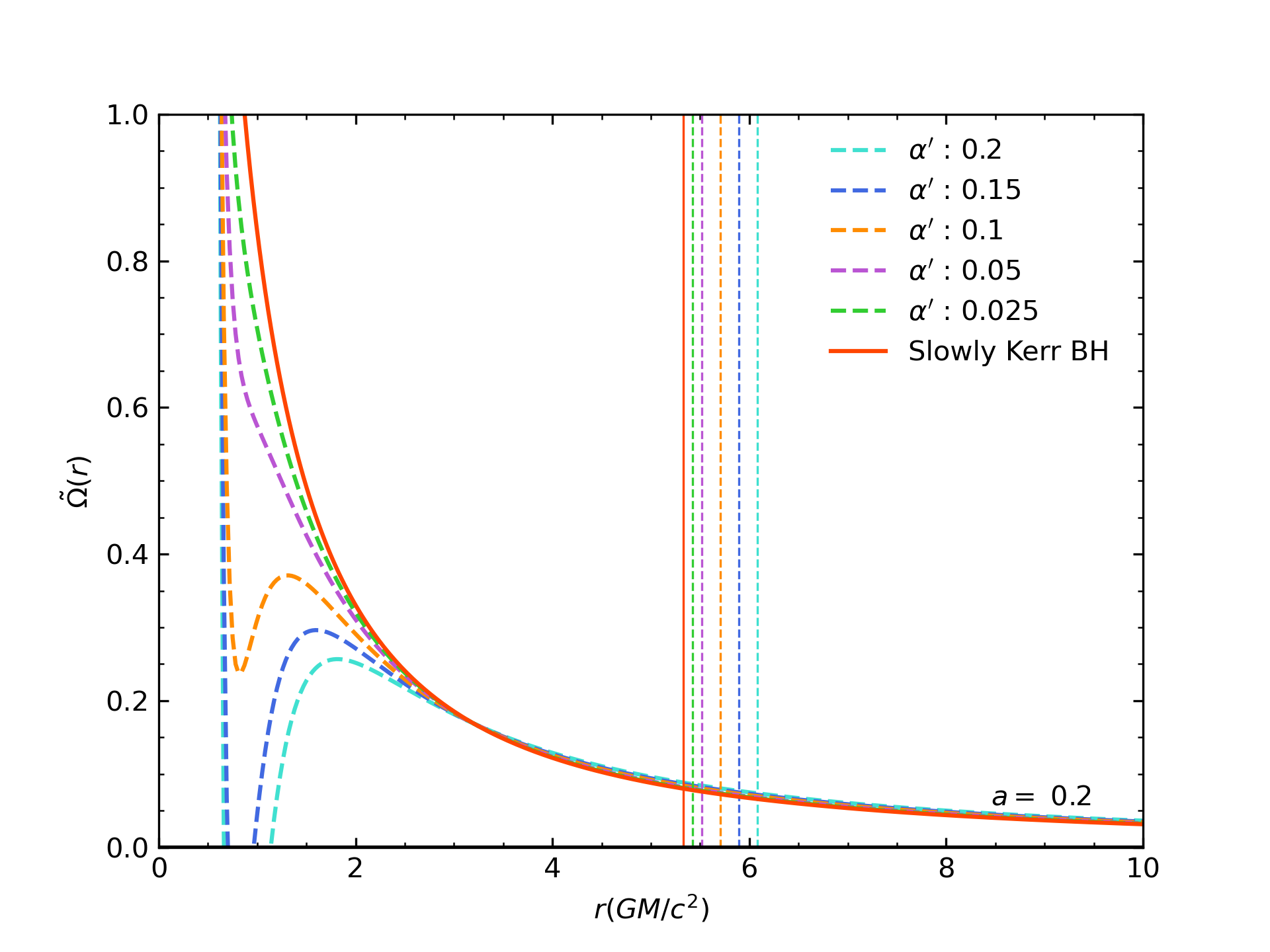}
        \includegraphics[scale=0.5]{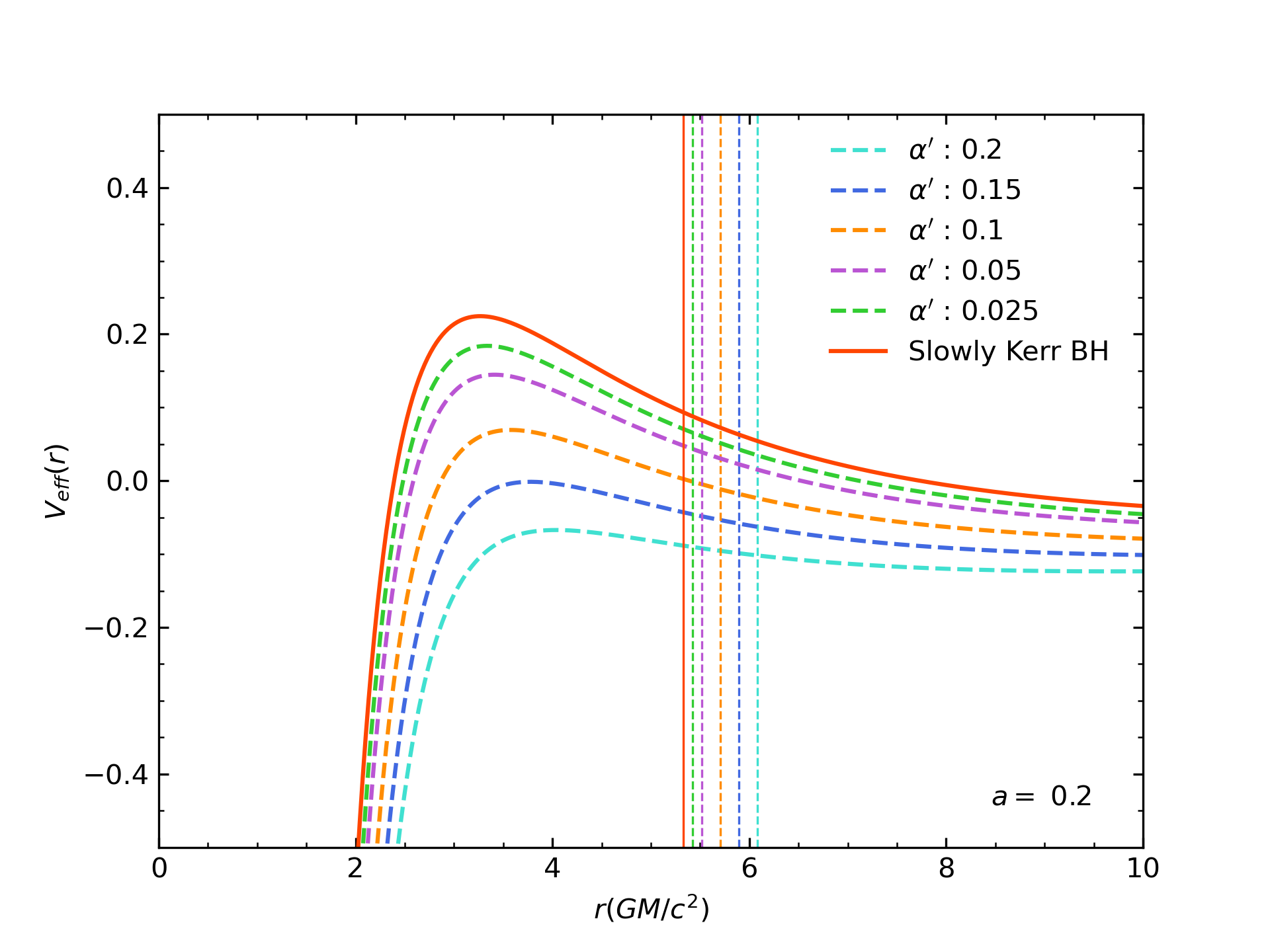}
        \caption{The upper-left panel shows the specific energy $\tilde{E}(r)$, the upper-right panel $\tilde{L}(r)$, the bottom-left panel $\Omega(r)$ and the bottom-right panel shows the effective potential for the values of $E=1$ and $L=4.5$, for the $\alpha'$-corrected BH with a mass of $M = 6.2 \times 10^{9} M_{\odot}$ and different values of $\alpha$ from $0$ to the upper limit $\sim 2$ in equation (\ref{upper-limit}).}
	\label{E-L-V}
\end{figure*}

Figure \ref{E-L-V} depicts the specific energy, angular momentum, Keplerian frequency and the effective potential from equations (\ref{Keplerian-frequencyfinal})-(\ref{effective-potential}) for different values of the coupling $\alpha$. It can be noticed that as the value of $\alpha$ increases and approaches its upper limit, clearer differences appear in the Keplerian frequency, energy and specific angular momentum. As a test particle approaches the event horizon, when looking at the plots of the energy and specific angular momentum for values of $\alpha$ larger than $0.05$ it seems that an unstable circular orbit emerges near the horizon within the ISCO, as well as a more intricate behaviour in the Keplerian frequency.

\subsection{Photon rings of slowly rotating $\alpha'$-corrected black hole}
We address now the problem of the separability of the Hamilton-Jacobi
equation for null geodesics, namely
\begin{align}
g^{\mu\nu}\frac{\partial S}{\partial x^{\mu}}\frac{\partial S}{\partial x^{\nu}} & =0\,.\label{HJ equation}
\end{align}
We assume a separable ansatz for the Jacobi characteristic function
\begin{align}
S & =-Et+L\phi+W(r)+P(\theta) \ .
\end{align}
Interestingly, equation (\ref{HJ equation}) is compatible with
the separable ansatz, leading to the following two equations 
\begin{align}
\left(\frac{\dd P}{\dd\theta}\right)^{2} & +\frac{\xi^{2}}{\sin^{2}\theta}-a^{2}\cos^{2}\theta-\xi^{2}-\eta=0\,,
\end{align}
and
\begin{align}
-\tilde{\Delta}\left(\frac{\dd W(r)}{\dd r}\right)^{2}+\frac{(\xi^{2}-4M^{2})r-2M\xi^{2}}{r(2M-r)^{2}}a^{2}+\nonumber \\
+\left(\frac{4\xi(4M^{2}+3Mr-3r^{2})}{3r^{2}(2M-r)^{2}}\alpha-\frac{4M\xi}{2M-r}\right)a+\nonumber \\
-\frac{2(12M^{3}+5M^{2}r+6r^{2}M-6r^{3})}{3(2M-r)^{2}M}\alpha+\nonumber \\
+\frac{(r-2M)\xi^{2}-r^{3}}{2M-r}-\eta & =0
\end{align}
where $\eta$ is the separation constant and we have defined $\xi\equiv\tilde{E}/L$
normalizing all the quantities respect to $\tilde{E}$. Without
loss of generality, we fix $\tilde{E}=1$, and define
\begin{align}
\tilde{\Delta} & \equiv r^{2}-2Mr+a^{2}+\frac{-40M^{2}+6Mr+6r^{2}}{3r^{2}}\alpha\,.
\end{align}
 From here we define the functions
\begin{align}
R(r) & \equiv\frac{\tilde{\Delta}^{2}}{\tilde{E}^{2}}\left(\frac{\dd W(r)}{\dd r}\right)^{2}\,,\\
\Theta(\theta) & \equiv\frac{1}{\tilde{E}^{2}}\left(\frac{\dd P(\theta)}{\dd\theta}\right)^{2}\,.
\end{align}
Following these definitions the Jacobi characteristic functions is
written as

\begin{align}
S & =-t+\xi\phi+\int_{r_{0}}^{r}\dd r\frac{\sqrt{R(r)}}{\tilde{\Delta}}+\int_{\theta_{0}}^{\theta}\dd\theta\sqrt{\Theta(\theta)}
\end{align}
and the first order equations for the non-trivial coordinates reduce to
\begin{align}
g_{rr}^{2}\dot{r}^{2} & =\frac{R(r)}{\tilde{\Delta}^{2}}\,,\\
g_{\theta\theta}^{2}\dot{\theta}^{2} & =\Theta(\theta)
\end{align}
Then it is possible to find conditions for the existence of circular
orbits without restricting to the equatorial plane, by imposing the
following two conditions for all $r$ 
\begin{align}
R(r) & =0\,,\\
\frac{\dd R(r)}{\dd r} & =0\,.
\end{align}
These equations can be solved in terms of the constants $\xi$ and $\eta$ giving two families
of solutions. Considering the branch that connects with the Kerr solution perturbatively, we have
\begin{align}
\xi & =a\frac{M+r}{M-r}+\frac{r(r^{2}-3Mr)}{a(M-r)}+\\
 & +\frac{a\alpha}{6r^{4}(2M-r)^{2}(M-r)M}(-240M^{7}+592M^{6}r+\nonumber \\
 & +62M^{5}r^{2}-347M^{4}r^{3}+44 M^{3}r^{4}+25M^{2}r^{5}+24Mr^{6}-12r^{7})+\nonumber \\
 & +\frac{\alpha}{a}\frac{1}{3r(M-r)^{2}(2M-r)M}(-336M^{5}+327M^{4}r+\nonumber \\
 & -66M^{3}r^{2}+16M^{2}r^{3}-18r^{4}M+3r^{5})\nonumber 
\end{align}
and
\begin{align}
\eta & =\frac{4Mr^{3}}{(M-r)^{2}}-\frac{(3M-r)^{2}r^{4}}{(M-r)^{2}a^{2}}+\frac{\alpha}{3r^{2}(M-r)^{3}(2M-r)^{2}}\times\nonumber \\
 & \times(-720M^{7}+2016M^{6}r-118M^{5}r^{2}-1619M^{4}r^{3}\nonumber \\
 & +781M^{3}r^{4}-47M^{2}r^{5}+21Mr^{6}-18r^{7})+\nonumber \\
 & -\frac{\alpha}{a^{2}}\frac{2r(r-3M)}{(r-2M)(r-M)^{3}}(r^{5}-6r^{4}M+\frac{16}{3}M^{2}r^{3}+\nonumber \\
 & -22M^{3}r^{2}+109M^{4}r-112M^{5}).
\end{align}
The impact parameters $x'$ and $y'$ defined as in \citet{bardeen1973black}, are the usual Cartesian coordinates of the image plane for a observer located at $r=r_0$ coordinate distance and with inclination angle of view $\theta = i$. In the case of $r_0 \rightarrow \infty$, the expression for the impact parameters are
\begin{eqnarray}
    x' &=& -\frac{\xi}{\sin i}, \\
    y' &=& \pm \sqrt{\Theta(i)},
\end{eqnarray}
where $\Theta(i) = a^2\cos^2 i - \xi^2 \cot i +\eta$.

\begin{figure*}
	\centering
	\includegraphics[scale=0.5]{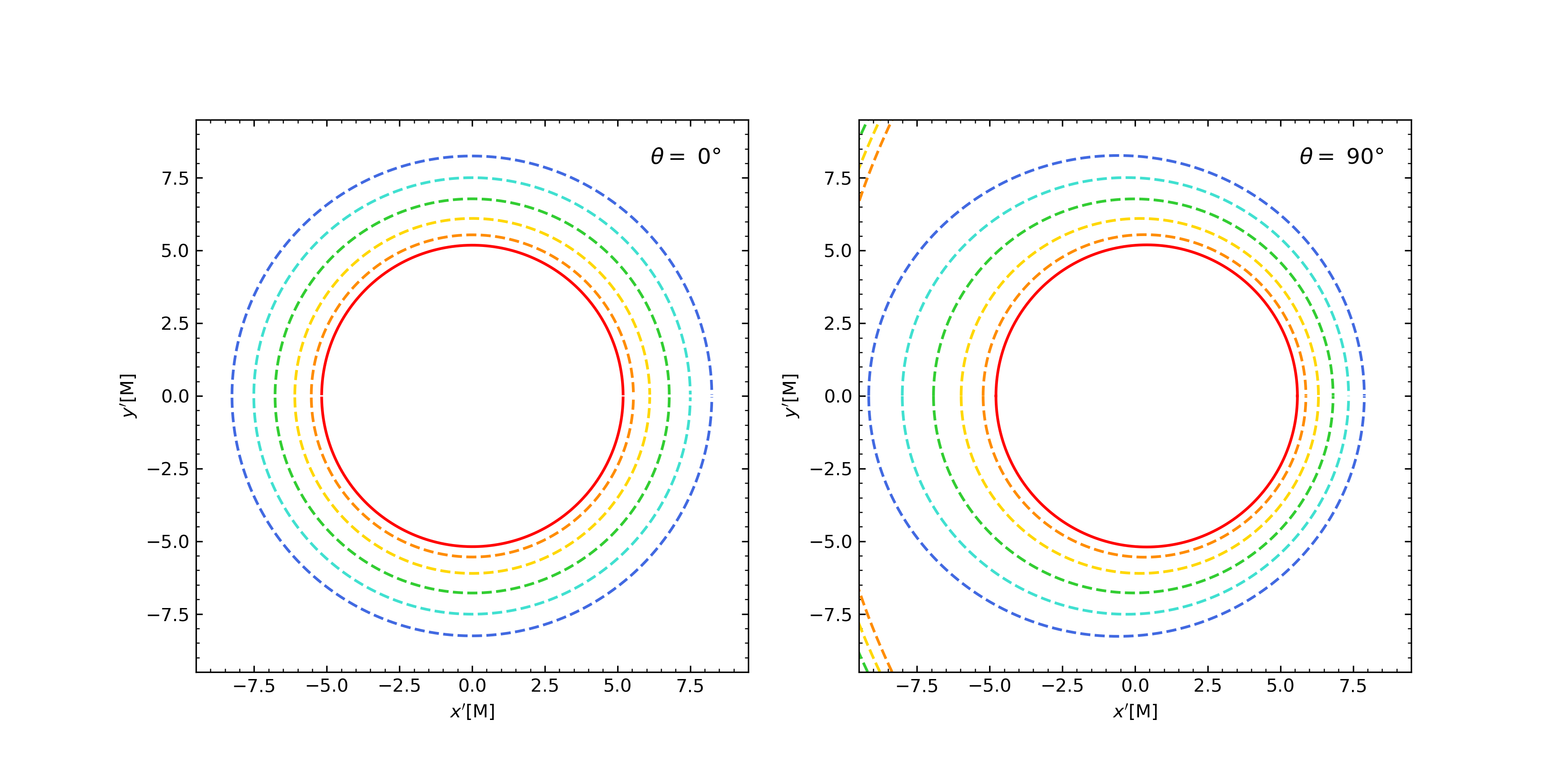}
        \caption{Photon rings around the $\alpha'$-corrected black holes at an inclination angle of $\theta = i = 0^{o}$ (left panel) and $i = 90^{o}$ (right panel), where the (\textit{red line}) represents a black hole with spin parameter $a=0.2 M$ in the Kerr approximation. The (\textit{orange dashed line} to \textit{blue dashed line}) correspond to black holes with the coupling $\alpha = [0.04,0.08,0.12,0.16,0.2]$, respectively, in units of the gravitational radius squared $r_g^2$.}
	\label{fig:Alpha_ring}
\end{figure*}

The shape of the photon rings in the usual Kerr metric has been already studied previously by several authors \citet{bardeen1973black,luminet1979image,takahashi2004shapes,beckwith2005extreme,amarilla2010null,johannsen2013photon,giribet2023sub}. Using the previous equations it is possible to obtain images of the photon rings in the slowly rotating $\alpha'$-corrected metric (see Figure \ref{fig:Alpha_ring}) in order to observe the effects of $\alpha$ in the shape of the photon ring and to compare it with the Kerr case. 

\section{Numerical model} \label{Accretion disk}
We describe now our numerical model, including the assumptions for the structure of the thin disk as well as the ray-tracing code RAPTOR I \citep{bronzwaer2018raptor}.

\subsection{Thin-disk model}\label{Thin-disk}

To analyze and obtain the shadow of the black hole, we assume a thin accretion disk, using the standard relativistic thin-disk model from \citet{novikov1973astrophysics}. This model describes a geometrically thin disk with steady and axially symmetric states of the matter around the black hole. The distribution of the matter starts at the location of the ISCO radius and extends to some very large radius $r_{out}$, forcing the particles to move in nearly circular geodesics in the equatorial plane along the disk.

From the equations of conservation of energy, rest mass, and angular momentum in the disk paticles \citep{novikov1973astrophysics}, the radiation flux emitted by the surface of the disk follows as

\begin{equation}\label{Flux}
    F(r) = \frac{\dot{M}}{4 \pi M^{2}} f_{disk}(r),
\end{equation}
where $\dot{M}$ is the mass accretion rate and
\begin{equation} \label{fdisk}
     f_{disk}(r) =  -\dfrac{d\Omega}{dr} \dfrac{M^{2}}{\sqrt{-g} (\tilde{E} - \Omega \tilde{L})^{2}} \int_{r_{in}}^{r} (\tilde{E}-\Omega \tilde{L} ) \dfrac{d \tilde{L}}{d\overline{r}} d\overline{r}
\end{equation}
is a dimensionless function of the radius, $\Omega$ is the Keplerian frequency, $\tilde{E}$ and $\tilde{L}$ are the specific energy and angular momentum, $r_{in}$ is the location of the ISCO radius and $g$ is the determinant of the metric in cylindrical coordinates $(t,r,z,\phi)$. In the case of SK we have $\sqrt{-g} = r$ and for the metric defined in Eq. (\ref{alpha-metric}), one has

\begin{equation}
    \sqrt{-g} = \sqrt{r^{2} - \alpha \left(\frac{32M^2 + 18Mr + 12r^2}{3Mr}\right)}.
\end{equation}

\begin{figure}[h!]
	\centering
	\includegraphics[scale=0.5]{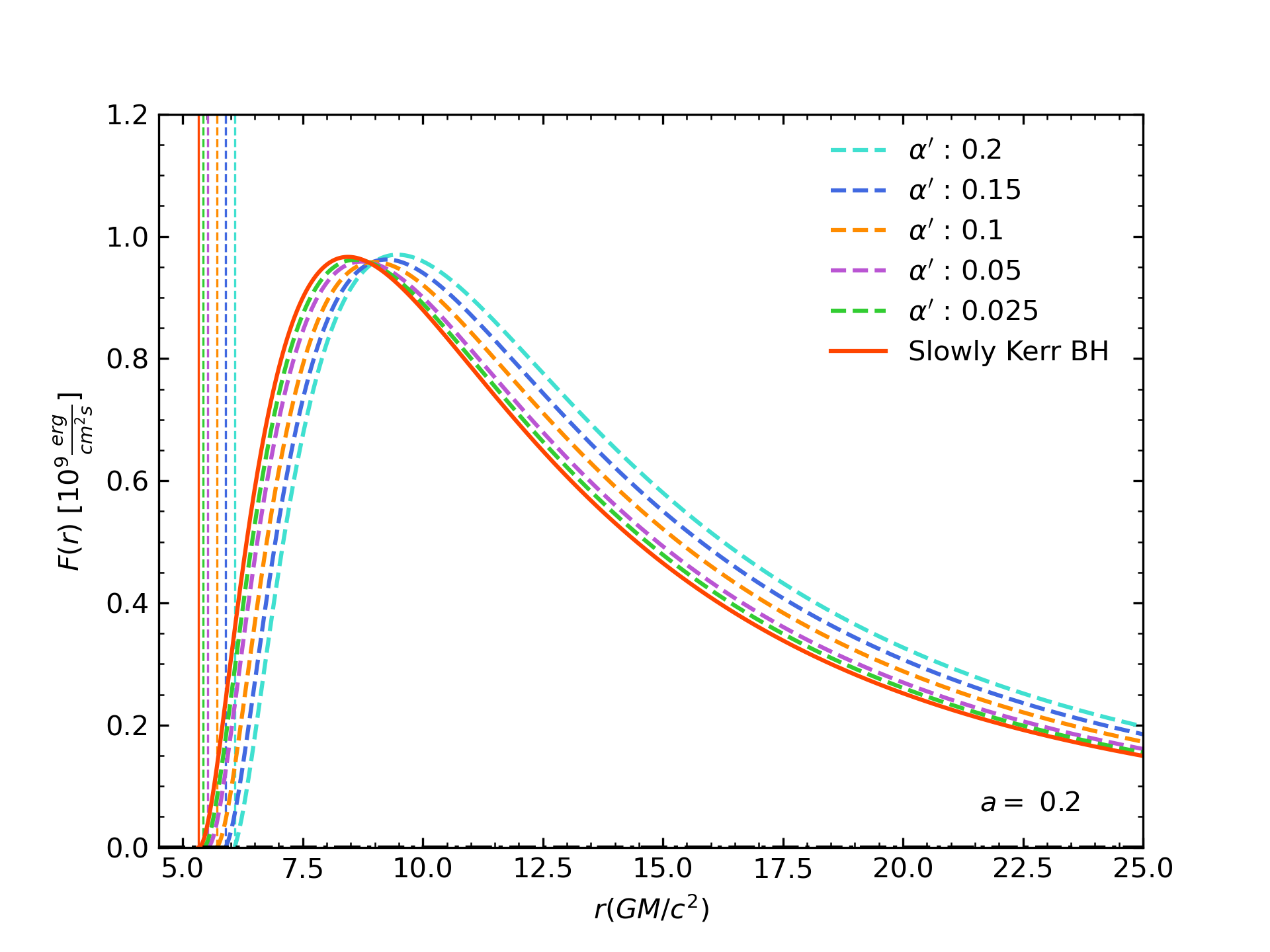}
        \caption{The energy flux of a BH with a rotation parameter $a=0.2M$, mass of $M_{BH} = 6.2 \times 10^9 M_{\odot}$ and mass accretion rate $\dot{M} = 1.172 \times 10^{23} [g/s]$ \cite{akiyama2019first} for different values of $\alpha$, where vertical lines represents the ISCO value for each $\alpha$.}
	\label{fig:Flux}
\end{figure}

Figure \ref{fig:Flux} shows the flux for this space-time compared to the slowly rotating Kerr case, where it can be seen that as $\alpha$ grows the flux moves slightly outwards from the black hole.

Since the radiative description of this model is defined by thermal equilibrium black-body emission due to its simplicity and effectiveness, the flux is related to the disk temperature according to the standard Stefan-Boltzmann law as
\begin{equation}
    F(r) = \sigma T_{eff}(r)^{4},
\end{equation}
where $\sigma$ is the Stefan-Blotzmann constant and $T_{eff}(r) =  f_{col}^{-1} T_{col}(r)$. Here the specific intensity can be defined as

\begin{equation}\label{Intensity-local}
    I_{\nu} = \frac{1}{f_{col}^{4}} B_{\nu}(\nu, T_{col}),
\end{equation}
where $f_{col}$ is the spectral hardening factor and 
\begin{equation}
    B_{\nu}(\nu,T_{col}) = \dfrac{2h\nu^{3}}{c^{2}} \dfrac{1}{\exp (h\nu/k_{B}T_{col}) - 1 }
\end{equation}
is the Planck distribution for a blackbody spectrum. Here $\nu$ is the frequency of emission and $h$ and $k_{B}$ are the Planck and Boltzmann constants, respectively. 

\subsection{Ray-tracing code RAPTOR I}
To obtain images of the BH shadow in the slowly rotating $\alpha'$ space-time, we employ a modified version of the open-source ray tracing code RAPTOR I \citep{bronzwaer2018raptor}, which is capable to work in arbitrary space-times. The code calculates the null geodesic around the BH and solves the relativistic radiative transfer equation along the geodesics. To use it for obtaining images of BHs in this specific space-time, we modified the metric of the code, the connection and also included the thin accretion disk model \citep{novikov1973astrophysics} with the specific changes of the different physical quantities such as the location of the ISCO, the velocity of the disk and the flux emitted according to this model as we discussed  in section~\ref{Thin-disk} for this space-time.

The code starts creating a virtual camera at a distant observer position  to define the initial conditions for the light rays; this is performed following the general approach from \citet{bambi2017black}.

Therefore, RAPTOR I integrates the geodesics along the path of the rays  which interacts with the structure of the disk that extends from the ISCO radius (\ref{ISCO-eq2}) to $r_{out}=1000 r_{g}$. In a final step the code employs the calculation of the radiative transfer equation at the observer's location as 
\begin{equation}
    I_{\nu, obs} = \overline{g}^{3} I_{\nu},
\end{equation}
where $I_{\nu}$ is the intensity emitted from Eq. (\ref{Intensity-local}) and $\overline{g}$ is the redshift factor
\begin{equation}
    \overline{g} = \frac{\nu_{obs}}{\nu}, 
\end{equation}
where $\nu_{obs}$ is the frequency at observer frame and 
\begin{equation}
    \nu = k^{\mu} u_{\mu}
\end{equation}
is the frequency in the plasma frame, $k^{\mu}$ is the contravariant wave vector and $u_{\mu}$ is the covariant four-velocity.

\section{Black hole simulations and shadow characterization}

In the following, we present our simulation results with an analysis of the diameter, displacement and asymmetry of the obtained images.

\subsection{Diameter, displacement and asymmetry}

To characterize the shadow of the different black holes in order to compare them, we employ the method used by \cite{johannsen2013photon,Agurto-Sepulveda:2023who} that consists in the study of the peaks of the intensity profile in each image of the black hole to calculate the displacement, diameter and asymmetry through the following expressions. The displacement is defined as
\begin{equation} \label{Displacement}
    D \equiv \frac{|x_{max} + x_{min}|}{2},
\end{equation}
where $x_{max}$ and $x_{min}$ are the locations of the two peaks in the horizontal intensity profile. Analogously, it is possible to define the displacement in the vertical direction. Then the average radius is defined by
\begin{equation}
    \langle \overline{R} \rangle \equiv \dfrac{1}{2 \pi} \int_{0}^{2\pi} \overline{R} d\alpha,
    \label{Radius}
\end{equation}
where $\overline{R} \equiv \sqrt{(x' - D_{x})^{2} + (y' - D_{y})^{2}}$. Thus the diameter is $L = 2 \langle \overline{R} \rangle$. 

Consequently, we get the following expression for the asymmetry: 
\begin{equation}
     A \equiv 2\sqrt{\dfrac{\int_{0}^{2\pi} \left(\overline{R} - \langle \overline{R} \rangle \right)^{2} d \alpha}{2\pi}}.
    \label{asymmetry}
\end{equation}

\subsection{Simulations}
\begin{figure*}
        \centering
	\includegraphics[scale=0.45]{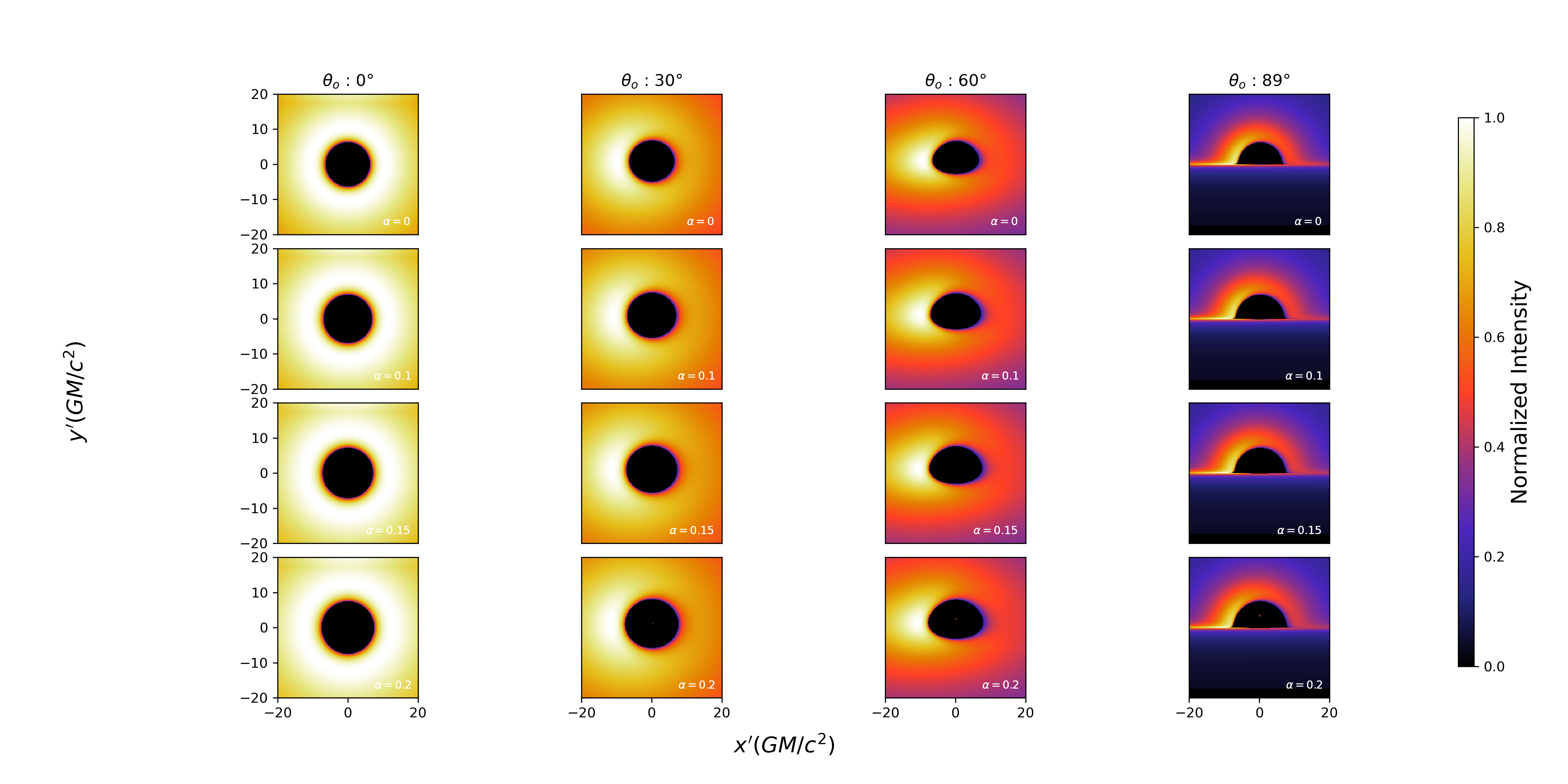}
	\caption{Simulation performed by Raptor I code of the Novikov-Thorne thin accretion disk model for a slowly rotating $\alpha'$-corrected SMBH with a mass of $M=6.2 \times 10^{9} M_{\odot}$, a mass accretion rate of $\dot{M} =1.172 \times 10^{23} [g/s]$, a spin parameter $a=0.2M$ and couplings $\alpha = [0.1,0.15,0.2]$, in units of the $r_g^2$.}
	\label{Simulations}
\end{figure*}

We performed simulations with RAPTOR I of an $\alpha'$-corrected BH that are shown in Figure \ref{Simulations}, with mass similar to M87*, for a thin accretion disk model. The settings of the simulations are shown in Table \ref{tabla1}.

\begin{table}[!h]
\centering
\caption{Setup for the simulation, values extracted from \citet{akiyama2019first}}
\begin{tabular}{lc}
\hline\hline\noalign{\smallskip}
\!\!Variable & \!\!\!\! Value  \\
\hline\noalign{\smallskip}
\!\!Mass  &  $6.2 \times 10^{9} $ $\mathrm{M_{\odot}}$\\
\!\!Distance & $16.9 $ $\mathrm{Mpc}$\\
\!\!$a$ & $0.2 \ M$  \\
\!\!$r_{\text{camera}}$ & $10^{4} \ \mathrm{r_{g}}$\\
\!\!Range for $(x^{\prime}$, $y^{\prime})$ & $[-20,20] \ \mathrm{r_{g}}$ \\
\!\!Resolution $(x,y)$ & $200$ $\mathrm{px}$ \\
\!\!Frequency & $230$ $\mathrm{GHz}$ \\
\!\!Inclination (°) & $[0,30,60,90]$ \\
\!\!$\dot{M}$ & $1.71 \times 10^{23} \mathrm{g/s}$ \\
\!\!$\alpha'$ & $[0, 0.1, 0.15, 0.2]$\\
\hline \hline
\end{tabular}
\label{tabla1}
\end{table}

To compare the shadow and disk structure of this BH with the usual Kerr BH in the same physical conditions, we performed a similar analysis as \citet{Agurto-Sepulveda:2023who}, calculating the intensity profile of each image, for different viewing angles from face-on to edge-on and then we obtained the values of the diameter, displacement and asymmetry of the shadow from equations (\ref{Displacement}-\ref{asymmetry}) for each simulation. 

\begin{figure*}
    \centering
    \includegraphics[scale=0.5]{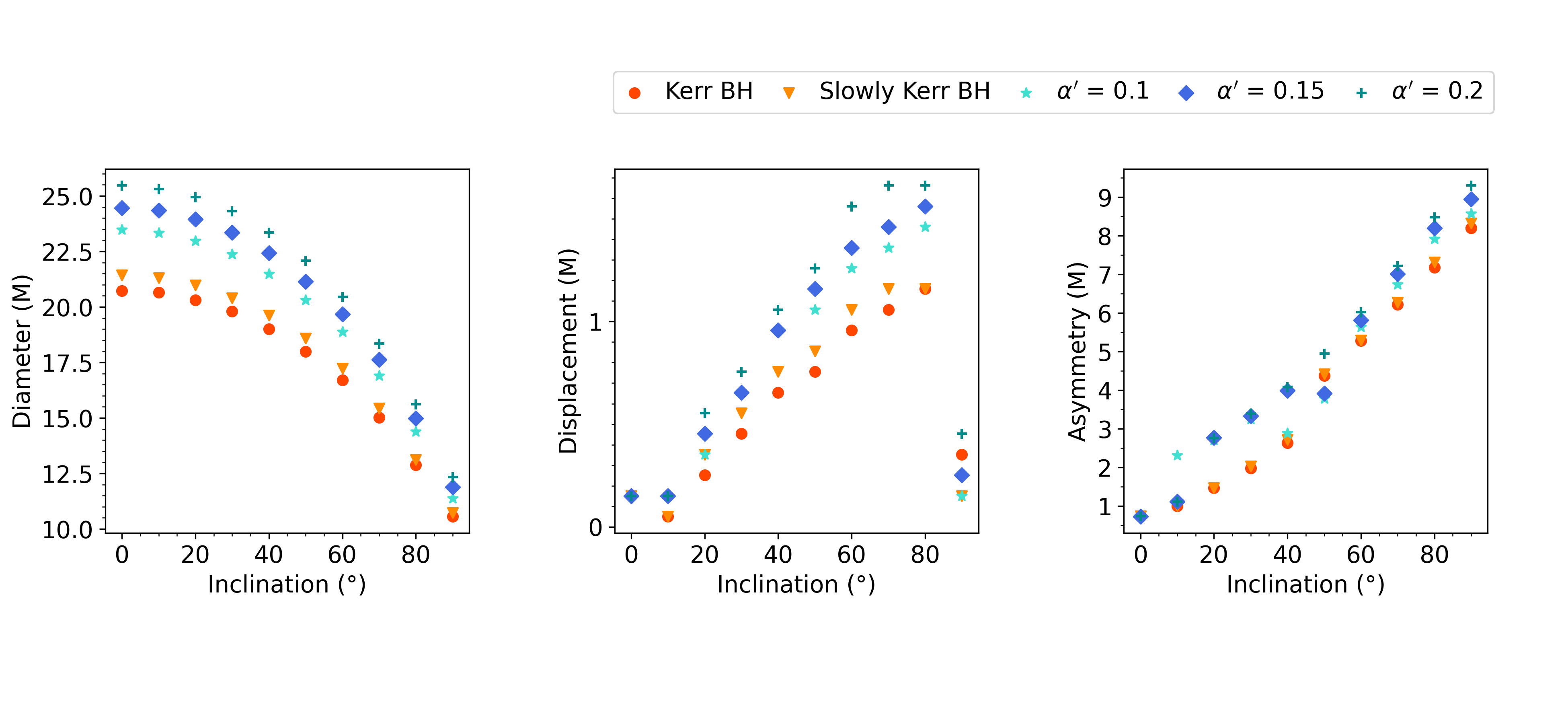}
    \caption{Comparison of the diameter, displacement and asymmetry values, for simulations of Kerr space-time (\textit{red dots}), slowly rotating Kerr approximation (\textit{orange triangles}) and $\alpha'$-corrected BH with different values of the coupling term $\alpha = [0.1,0.15,0.2] \ r_{g}^2$ (\textit{cyan stars, blue diamonds and darkcyan plus sign}, respectively).}
    \label{Deviations_comparison}
\end{figure*}

A comparison is shown in Figure \ref{Deviations_comparison}, where one can see that an increasing value of $\alpha'$ leads to an increase of the diameter of the image  compared to Kerr while it decreases when going from a face-on to an edge-on viewing angle. Here, it is also important to notice that the Kerr case and the slowly rotating Kerr case also differ in the diameter for a few values of inclination. In the case of the displacement, we can see a similar situation, where the displacement of the shadow increases with the value of $\alpha'$, and decreases abruptly for an edge-on viewing angle. Here, the difference  between the Kerr case and the $\alpha'$-corrected case is also clearly visible. Finally, the asymmetry values do not differ as much as their predecessors and are quite close to the Kerr case, yet for increasing $\alpha'$  differences can be noticed for the first lower half of the viewing angles and stay close to Kerr's case for the top half.

\subsection{Effects of $\alpha'$-corrected terms in the shadow}
One way to observe possible signals of the deviations from the Kerr shadow due to the $\alpha'$-corrected terms, is setting the structure of the disk from a given value of ISCO radius $r_{isco}$ to a big outer radius $r_{out}$ for the two different space-times in the simulations. In the static case, the radius of the shadows of the Callan-Myers-Perry black holes \cite{Callan:1988hs} was obtained in \cite{Moura:2021eln}.

As we have seen before, increasing the value of $\alpha'$ correspond to an increase in the diameter and displacement of the BH shadow. Since $\alpha'$ stands for the inverse of the tension of the string, it must be positive. Therefore, to have a same value for the ISCO radius with different values of $\alpha'$ and spin parameter $a$ we proceed as follows: we fix the ISCO radius as $r_{isco} =  7.31 r_{g}$, which in Kerr metric yields a value of spin parameter $a=-0.42M$, while in the slowly rotating $\alpha'$ metric yields a spin parameter $a= -0.3M$ and the coupling value of $\alpha = 0.189 \ r_{g}^2$.

\begin{figure}
    \centering
    \includegraphics[scale=0.4]{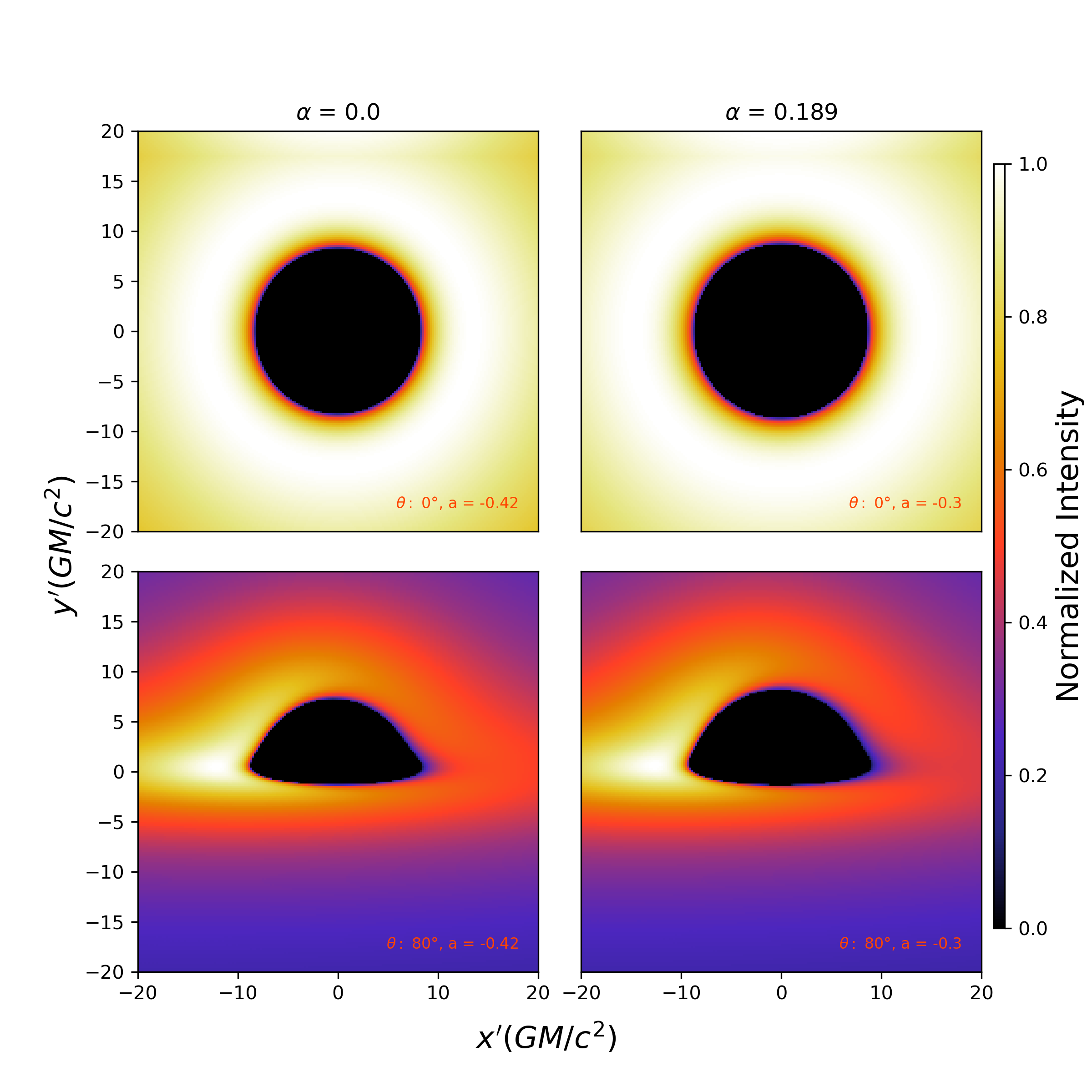}
    \caption{Comparison between two simulations with the similar configuration of Table \ref{tabla1} for a BH in a Kerr space-time (\textit{left panel}) with a spin parameter $a = -0.42M$ and inclination angle of view $\theta =  [0,80]$ degrees and a slowly rotating $\alpha'$-corrected BH (\textit{right panel}) with different values of the coupling term $\alpha = 0.189 \ r_{g}^2$ and spin parameter $a=-0.3M$.}
    \label{Kerr_vs_alpha}
\end{figure}

Figure \ref{Kerr_vs_alpha} shows the difference between the two images of a Kerr black hole (\textit{left panel}) and the slowly rotating $\alpha'$-corrected black hole (\textit{right panel}) with the same location of the ISCO corresponding to the previously described setting. A careful look at the images shows a slight difference in the shadow size, being larger in the case of $\alpha'-$corrected space-time, as well as a slight difference in the intensity of the images.

\begin{figure*}
    \centering
    \includegraphics[scale=0.5]{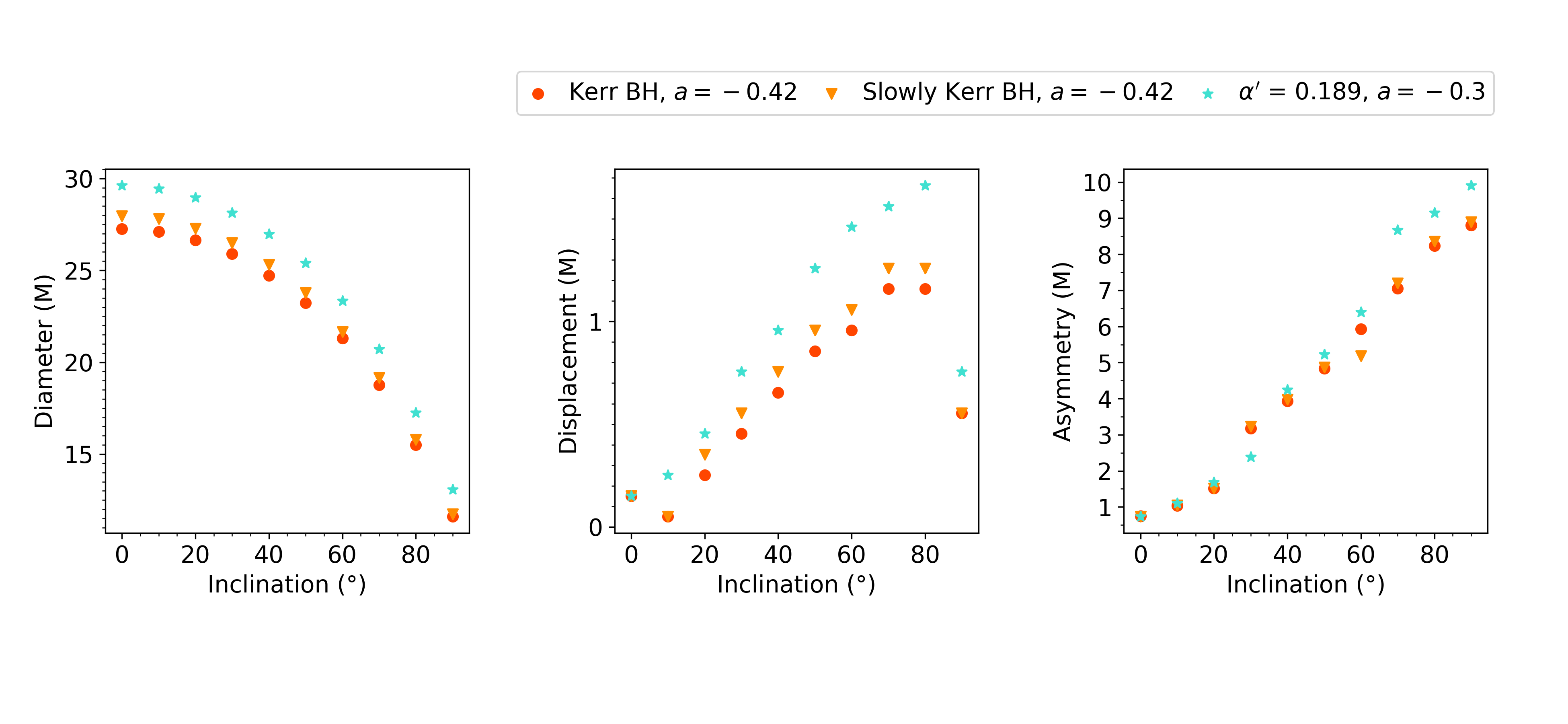}
    \caption{Comparison of the diameter, displacement and asymmetry values, for simulations of BHs with the same ISCO radius in the Kerr space-time (\textit{red dots}) and slowly rotating Kerr approximation (\textit{orange triangles}) with a retrograde spin parameter $a=-0.42 M$ and $\alpha'$-corrected BH with a spin parameter $a = -0.3 M$ and $\alpha' = 0.189$.}
    \label{Comparison_kerr_alpha}
\end{figure*}

In Figure \ref{Comparison_kerr_alpha} we show the differences in the diameter, displacement and asymmetry values for different angles of view for the same configuration, including also a comparison with the slowly rotating Kerr spacetime. There, it is possible to notice a greater difference between the images. The $\alpha'$-corrected BH shows a bigger diameter compared to the Kerr BH, for an angle of view near face-on, while the diameters get closer as one approaches the edge angle. In the case of the displacement, it is clearly seen that for increasing angle of view the displacement also increase for the $\alpha'$-corrected BH, when compared to the Kerr BH. In the case of asymmetry, the difference between the two images seems to be less clear, a slight difference may be visible for an angle of view closer to the edge-on, but still the values seem to be close to each other.

Table \ref{tabla2} shows several simulations with the same approach, fixing the ISCO radius, which were performed to observe the effect of $\alpha'$ for different values of spin $a$ and the same angle of view of $75^{o}$.

\begin{table}[htbp]
\centering
\caption{Comparison between the diameter, displacement and asymmetry for a set of simulations with the same ISCO radius and different values of the spin parameter $a$ and the coupling $\alpha'$.}
\begin{tabular}{crrrrrc}
\hline\hline\noalign{\smallskip}
\!\!$r_{isco} \ [\mathrm{r_{g}}]$ & \!\! $a \ [\mathrm{M}]$ & \!\!$\alpha^{\prime} [\mathrm{r_{g}^2}]$ & \!\! $L \ [\mathrm{r_{g}}]$ & \!\! $D \ [\mathrm{r_{g}}]$ & \!\! $A \ [-]$  & \!\! $\theta \ [{}^{o}$]  \\
\hline\noalign{\smallskip}
&  $-0.05$ & $0$ & $15.389$  & $1.158 $ &$7.121$& \\
$6.162 $  &$0.05$ & $0.102$ & $16.481 $ & $1.461 $ &$7.555$& $75$\\
&  $0.1$ & $0.145$ & $16.816 $ & $1.561$ &$7.753$& \\
&  $0.15$ & $0.185$ & $17.105$ & $1.561$ &$7.857$&\\ \hline \hline
 &  $-0.1$ & $0$ & $15.642$ & $1.057$ &$7.167$&\\
$6.322$&  $-0.05$ & $0.056$ & $16.375$ & $1.360$ &$7.472$& $75$\\
&  $0.05$ & $0.152$ & $17.146$ & $1.561$ &$7.814$& \\
&  $0.1$ & $0.193$ & $16.415$ & $1.662$ &$8.581$& \\  \hline \hline
& $-0.15$ & $0$ & $15.883$ & $1.158$ &$7.258$& \\
$6.481$ &$-0.1 $ & $0.059$ & $16.705$ &$1.360$ & $7.583$& $75$\\
&  $-0.05$ & $0.112$ & $17.151$ & $1.461 $ &$7.774$&\\
&  $0.05$  &  $0.201$ & $17.647$ & $5.088 $ &$9.524$&\\
\hline \hline
& $-0.2$ & $0$ & $16.143$ & $1.158$ &$7.358$& \\
$6.639$ &$-0.15 $ & $0.063$ & $17.031$ &$1.360$ & $7.680$& $75$\\
&  $-0.1$ & $0.119$ & $17.419$ & $1.461 $ &$7.855$&\\
&  $-0.05$  &  $0.168$ & $17.733$ & $1.662$ &$8.035$&\\
\hline \hline
& $-0.3$ & $0$ & $16.659$ & $1.158$ &$7.502$& \\
$6.949$ &$-0.25 $ & $0.074$ & $17.590$ &$1.360$ & $7.849$& $75$\\
&  $-0.2$ & $0.136$ & $18.039$ & $1.662 $ &$8.054$&\\
&  $-0.15$  &  $0.189$ & $18.459$ & $1.662$ &$8.268$&\\
\hline \hline
& $-0.4$ & $0$ & $17.153$ & $1.158$ &$7.639$& \\
$7.254$ &$-0.3$ & $0.160$ & $18.773$ &$1.561$ & $8.268$& $75$\\
\hline \hline
\end{tabular}
\label{tabla2}
\end{table}
Our results indicate that there are particular observable quantities, for which the differences on the BH image, for a given ISCO, are enhanced and therefore one could look for such observables to obtain a greater difference, when looking for qualitative signals beyond GR, in the context of the $\alpha'$-corrected black holes.

\section{Conclusions and discussions}
Summarizing, the solution that we presented in this paper is a extension of our slowly rotating solution described in \cite{Agurto-Sepulveda:2022vvf}, including the logarithmic branch from the static solution. We presented the conserved charges, the Hawking temperature and the entropy for this space-time in the particular case of $b=0$. Then, we studied several properties of the metric, first we assign a Petrov D type to the solution, we found a curvature singularity inside the event horizon, the location of the event horizon and ergosphere, as well as an upper limit for $\alpha$, all the quantities related to the Kerr metric. We then move on to the study of circular geodesics motion of test particles where we find that the circular orbits around this black hole are different from the Kerr orbit, presenting an unstable circular orbit close to the horizon. We also find that our solution is separable under the Hamilton-Jacobi equations, allowing us to find expressions for the photon rings where it is possible to see that as $\alpha$ grows, the ring diameter grows as well. Finally, we move on to the study of black hole shadowing by means of numerical ray tracing simulations for a thin accretion disc,  and then a post-analysis is carried out. Determining that increasing values of $\alpha$ increase the diameter of the shadow and subtly affect the shadow offset and asymmetry. This is evident from simulations where the accretion disc has the same starting point, i.e. the same ISCO radius. 

\begin{acknowledgments}
We thank B. Bandyopadhyay, A. Cisterna and G. Giribet for enlightening comments. FAS and DRGS gratefully acknowledge support via the ANID BASAL project  FB210003 and via Fondecyt Regular (project code 1201280). DRGS thanks for funding via ANID QUIMAL220002 and via the  Alexander von Humboldt - Foundation, Bonn, Germany. JO is partially supported by FONDECYT Regular 1221504. MO is partially funded by Beca ANID de Doctorado grant 21222264.\end{acknowledgments}

\nocite{*}

\bibliography{PRDV2}

\end{document}